\documentclass[numberedappendix,natbib209]{emulateapj}
\usepackage{longtable}
\usepackage{natbib}

\begin{document}

%\title{The Phase Space Locations and Stellar Populations of Poststarburst Galaxies in Clusters at \lowercase{z} $\sim$ 1: Evidence for a Rapid Environmental Quenching Timescale}
%\title{The Phase Space and Stellar Populations of Cluster Galaxies at \lowercase{z} $\sim$ 1: Simultaneous Evidence for Rapid Environmental Quenching Timescales in Cluster Cores}
\title{The Phase Space and Stellar Populations of Cluster Galaxies at \lowercase{z} $\sim$ 1: Simultaneous Constraints on the Location and Timescale of Satellite Quenching\altaffilmark{1}}

%\author{Adam Muzzin\altaffilmark{2}}
\author{Adam Muzzin\altaffilmark{2}, R. F. J. van der Burg\altaffilmark{2}, Sean L. McGee\altaffilmark{2}, Michael Balogh\altaffilmark{2,3}, Marijn Franx\altaffilmark{2}, Henk Hoekstra\altaffilmark{2}, Michael J. Hudson\altaffilmark{3,4}, Allison Noble\altaffilmark{5}, Dan S. Taranu\altaffilmark{5}, Tracy Webb\altaffilmark{6}, Gillian Wilson\altaffilmark{7}, H. K. C. Yee\altaffilmark{5}}

\altaffiltext{1}{Based on observations obtained at the Gemini Observatory, which is operated by the Association of Universities for Research in Astronomy, Inc., under a cooperative agreement with the NSF on behalf of the Gemini partnership: the National Science Foundation (United States), the Science and Technology Facilities Council (United Kingdom), the National Research Council (Canada), CONICYT (Chile), the Australian Research Council (Australia), Minist\'{e}rio da Ci\^{e}ncia e Tecnologia (Brazil) and Ministerio de Ciencia, Tecnolog\'{i}a e Innovaci\'{o}n Productiva (Argentina).}
\altaffiltext{2}{Leiden Observatory, Leiden University, PO Box 9513, 2300 RA Leiden, The Netherlands}
\altaffiltext{3}{Department of Physics and Astronomy, University of Waterloo, Waterloo, Ontario N2L 3G1, Canada}
\altaffiltext{4}{Perimeter Institute for Theoretical Physics, Waterloo, Ontario, Canada}
\altaffiltext{5}{Department. of Astronomy \& Astrophysics, University
  of Toronto, 50 St. George St., Toronto, Ontario, Canada, M5S 3H4}
\altaffiltext{6}{Department of Physics, McGill University, MontrŽal, QC, Canada}
\altaffiltext{7}{Department of Physics and Astronomy, University of California, Riverside, CA 92521}

%\altaffiltext{3}{Department of Physics and Astronomy, Tufts University, Medford, MA 06520, USA}
%\altaffiltext{4}{Physics and Astronomy Department, University of Missouri, Columbia, MO 65211}
%\altaffiltext{5}{Institut d'Astrophysique de Paris, UMR7095 CNRS, Universit\'{e} Pierre et Marie Curie, 98 bis Boulevard Arago, 75014 Paris, France}
%\altaffiltext{6}{Dark Cosmology Centre, Niels Bohr Institute, University of Copenhagen, Juliane Maries Vej 30, 2100 Copenhagen, Denmark}
%\altaffiltext{7}{SUPA, Institute for Astronomy, University of Edinburgh, Royal Observatory, Edinburgh EH9 3HJ, UK}
%\altaffiltext{8}{Laboratoire d'Astrophysique de Marseille, CNRS and Aix-Marseille Universit\'{e}, 38 rue Fr\'{e}d\'{e}ric Joliot-Curie, 13388 Marseille, Cedex 13, France}
%\altaffiltext{9}{European Southern Observatory, Alonso de C\'{o}rdova 3107, Casilla 19001, Vitacura, Santiago, Chile} 
\begin{abstract}
We investigate the velocity vs.~position phase space of $z \sim$ 1 cluster galaxies using a set of 424 spectroscopic redshifts in 9 clusters drawn from the GCLASS survey.  Dividing the galaxy population into three categories: quiescent, star-forming, and poststarburst, we find that these populations have distinct distributions in phase space.  Most striking are the poststarburst galaxies, which are commonly found at small clustercentric radii with high clustercentric velocities, and appear to trace a coherent ``ring" in phase space.  Using several zoom simulations of clusters we show that the coherent distribution of the poststarbursts can be reasonably well-reproduced using a simple quenching scenario.  Specifically, the phase space is best reproduced if satellite quenching occurs on a rapid timescale (0.1 $<$ $\tau_{Q}$ $<$ 0.5 Gyr) after galaxies make their first passage of R $\sim$  0.5R$_{200}$, a process that takes a total time of $\sim$ 1 Gyr after first infall.  The poststarburst phase space is not well-reproduced using long quenching timescales ($\tau_{Q}$ $>$ 0.5), or by quenching galaxies at larger radii (R $\sim$ R$_{200}$).   We compare this quenching timescale to the timescale implied by the stellar populations of the poststarburst galaxies and find that the poststarburst spectra are well-fit by a rapid quenching ($\tau_{Q}$ = 0.4$^{+0.3}_{-0.4}$ Gyr) of a typical star-forming galaxy.  The similarity between the quenching timescales derived from these independent indicators is a strong consistency check of the quenching model.  Given that the model implies satellite quenching is rapid, and occurs well within R$_{200}$, this would suggest that ram-pressure stripping of either the hot or cold gas component of galaxies are the most plausible candidates for the physical mechanism.  The high cold gas consumption rates at $z \sim$ 1 make it difficult to determine if hot or cold gas stripping is dominant; however, measurements of the redshift evolution of the satellite quenching timescale and location may be capable of distinguishing between the two.

\end{abstract}
\keywords{galaxies: clusters: general -- galaxies: evolution -- galaxies: formation -- galaxies: high-redshift}
\section{Introduction}
\indent
It is well known that galaxies in high-density environments such as galaxy groups and clusters (i.e., satellite galaxies) exhibit a higher fraction of quiescent galaxies at a fixed stellar mass than more isolated ``field" galaxies (i.e., central galaxies).  This is true both in the local universe \citep[e.g.,][]{Baldry2006,vandenbosch2008,Peng2010,Wetzel2012,Rasmussen2012,Haines2013}, and at $z \sim$ 1 \citep[e.g.,][]{Patel2009,Cooper2010,Sobral2011,Muzzin2012,Raichoor2012,vanderburg2013,Mok2013,Woo2013,Nantais2013,Kovac2014}.  While the correlation between galaxy quiescence and environment is well-established, and heuristic models that can explain the quenching rates and timescales exist \citep[e.g.,][]{Peng2010,Wetzel2012}, at present there is little direct observational evidence linking the satellite quenching mechanism to a specific physical process that occurs in clusters/groups such as ram-pressure stripping of cold gas \citep[e.g.,][]{Gunn1972} or hot gas \cite[e.g, ``strangulation",][]{Larson1980,Balogh1999}, mergers, or harassment \citep[e.g.,][]{Moore1996}.  One way to make further progress in identifying the dominant satellite quenching mechanism will be to better constrain both the timescale over which quenching occurs, and its location within the cluster/group \citep[e.g.,][]{Treu2003,Moran2007}.
\newline\indent
Semi-analytic models are physically motivated and recent works have argued that in order to properly reproduce the fraction of quiescent galaxies as a function of clustercentric radius, long quenching timescales, of order 3 - 7 Gyr, are necessary \citep[e.g.,][]{Weinmann2010,Mcgee2011,Delucia2012}.  Taken at face value, these long timescales are difficult to reconcile with observations, where a weak dependence of the specific star formation rates (SSFRs) of star-forming galaxies on environment is found \citep[e.g.,][]{Kauffmann2004,Peng2010, Vulcani2010,Muzzin2012,Wetzel2013}.  Indeed, the observations suggest a rapid quenching timescale \citep[e.g.,][although see Taranu et al.~2012 for evidence of longer timescales]{Muzzin2012,Wetzel2013,Mok2013}, and this hypothesis is supported by an abundance of poststarburst galaxies found in clusters at higher redshifts \citep[e.g.,][]{Poggianti2004,Tran2007,Poggianti2009a,Balogh2011,Muzzin2012,Mok2013,Wu2013}.  Interestingly, hydrodynamical simulations predict much faster quenching timescales than semi-analytic models \citep[e.g.,][]{McCarthy2008,Bahe2013,Cen2014}, and similar to the observations, they do not see the strong dependence of SSFR on environment.  
It has been argued by \cite{Wetzel2013} that one way to reconcile the long quenching timescales required by semi-analytic models with the short quenching timescales required by observations is to have a ``delayed-then-rapid" quenching, where galaxies experience a $\sim$ 2 -- 4 Gyr delay after infall into a cluster where they behave as normal star-forming galaxies before quenching in $<$ 1 Gyr.  
\newline\indent
While some progress towards defining the timescale of satellite quenching is being made, there are still few observational constraints on where within the cluster/group the process begins, and this is key information for identifying the physical processes involved \citep[e.g.,][]{Treu2003}.  Our previous work on $z \sim$ 1 clusters showed that poststarbursts are more common in the cluster core than in the outskirts \citep{Muzzin2012}; however, most previous studies simply compared the poststarburst fraction between cluster and field.  A better way of identifying populations within the cluster is to use the velocity vs. position phase space of clusters \citep[e.g.,][]{Biviano2002,Mahajan2011,Haines2013,Oman2013}.  Recently \cite{Noble2013} performed such an analysis on the most massive cluster in the GCLASS sample, and showed that this combined space is a more effective way of identifying sub-populations within the cluster than simply using clustercentric radius.  In particular, they showed that while properties such as SSFR and D$_{n}$(4000) show little dependence on clustercentric radius, once galaxies are separated in phase space there is a dependence of those properties as a function of environment.
\newline\indent
In this paper we continue the phase space analysis approach of \cite{Noble2013} using the full GCLASS sample and examine the location of cluster galaxies at $z \sim$ 1 in phase space.  In particular, we focus on the poststarburst population in order to try and identify the satellite quenching timescale and location at $z \sim$ 1.  Throughout the paper we assume a cosmology with H$_{0}$ = 70 km s$^{-1}$ Mpc$^{-1}$ and $\Omega_{m}$ = 0.3, $\Omega_{\Lambda}$ = 0.7.
%correlations, well-quantified, models, physics still poorly known.  Constraining timescales and quenching locations could help.  Disagreement between data and simulation quenching.  %Poststarbursts are key to this.  We (and others) have found a lot of these in the clusters, suggesting they are linked to the environmental quenching process.  While we see a rise in %the clusters, it is unclear where they are.  Phase space is useful (e.g., biviano).  Recently, Noble showed that dividing clusters into phase space was useful for figuring out what %happened.   Here we apply this to the full GCLASS sample and look at the poststarbursts as clues to the environmental quenching timescale and location.
%TBD: Read 1. Noble, 2. Wetzel, 3. Haines and 4. Mahjan+2011 and 5. Treu+2004, 6.McGee, 7.Biviano
%Papers to cite: Mok, McGee, De Lucia, McCarthy
\section{Dataset}
Our analysis is based on a spectroscopic sample of 424 cluster galaxies in 9 clusters at $z \sim$ 1 from the GCLASS survey \citep[see][]{Muzzin2012}.  The GCLASS clusters were selected from the 42 deg$^{2}$ SpARCS survey \citep{Muzzin2009a,Wilson2009} using an optical/IR adaptation of the red-sequence method \citep{Gladders2000} that is discussed in \cite{Muzzin2008}.  The clusters have halo masses between M$_{200}$ $\sim$ (1.0 - 20.0) $\times$ 10$^{14}$ M$_{\odot}$ \citep[][G. Wilson et al., in preparation]{vanderburg2014} which have been inferred from the cluster line-of-sight velocity dispersion ($\sigma_{v}$).  
\newline\indent
We classify each galaxy in the spectroscopic sample as either star-forming, quiescent, or poststarburst\footnote{In this paper, following convention, we refer to these galaxies as "poststarbursts". In fact, we will show later that these galaxies are well-fit by rapidly-truncated star-formation with no secondary "burst".  They could also be considered as ``recently quenched" galaxies, but we keep the poststarburst designation for consistency with previous work.} using the [OII] emission line and D$_{n}$(4000) as diagnostics.  Star-forming galaxies are classified as those galaxies with detected [OII] emission, where the detection limit is $\sim$ 1 - 3\AA~equivalent width (EW), depending on signal-to-noise.  Quiescent galaxies are defined as those without detected [OII] emission.  Similar to \cite{Muzzin2012}, poststarburst galaxies are defined as the subset of quiescent galaxies that have D$_{n}$(4000) $<$ 1.45 (i.e., those that are quiescent but have young stellar populations).  This is somewhat different than the more typical EW(H$_{\delta}$) $>$ 3 -- 5\AA~definition used for K+A galaxies in many studies \citep[e.g.,][]{Dressler1999,Balogh1999,Poggianti2009a}; however, as discussed in \cite{Muzzin2012}, H$_{\delta}$ is a weak line and is difficult to measure consistently in all spectra at $z \sim$ 1, and a weak D$_{n}$(4000) serves as a good proxy for strong H$_{\delta}$ in galaxies without [OII] emission.  The average stacked spectrum of our poststarburst definition does have EW(H$_{\delta}$) $>$ 5\AA~\citep[][see also $\S$ 5]{Muzzin2012}, and strong Balmer lines, so on average our selection of poststarbursts is comparable to the K+A selection criteria.  
\newline\indent
We note that the poststarbust classification may not be 100\% complete for all galaxies with strongly truncated star formation.  One population that would be missed are old galaxies that experienced a recent rejuvenation of star formation and then a subsequent truncation of that star formation.  If the total stellar mass formed in that event was modest, they will have D$_{n}$(4000) $>$ 1.45 and remained classified as quiescent.  It is unclear if such a population exists in clusters at $z \sim$ 1; however, if so, then it would be absent from the current sample.
\begin{figure*}
\centering
\includegraphics[width=0.45\textwidth, angle =90 ]{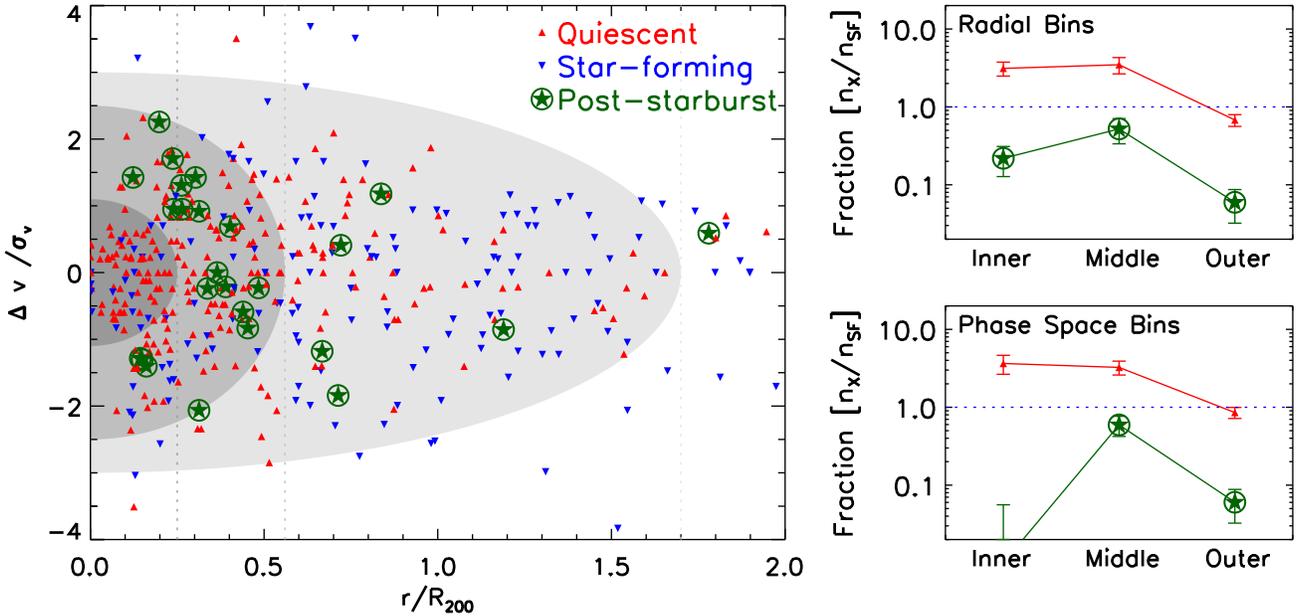}
%\plotone{velocityposition3.eps}
\caption{\footnotesize Left panel: The velocity vs. clustercentric radius phase space of galaxies in the 9 GCLASS clusters.  The velocities are in units relative to the individual cluster velocity dispersions and the radii are relative to the position of the brightest cluster galaxy scaled by the R$_{200}$ of the cluster.  The shaded regions are arbitrarily defined but are indicative of increasing time since infall (see text).  Quiescent galaxies (red triangles), star forming galaxies (blue triangles) and poststarburst galaxies (green stars) all occupy distinct locations in phase space.  Right panels: The ratio of quiescent and poststarburst galaxies compared to star-forming galaxies separated into the three radial bins marked by the dotted lines (top panel), and the three phase space bins marked by the shaded regions (bottom panel).  The error bars are 1$\sigma$ Poisson errors.  Poststarburst galaxies are distributed fairly uniformly in the cluster by radius (top panel), with a peak in the middle bin; however, in phase space they are most prevalent in the middle bin and completely absent in the inner bin (bottom panel).}
\end{figure*}
\section{Galaxies in the Cluster Phase Space}
In the left panel of Figure 1 we plot the velocity vs.~projected clustercentric radius phase space (hereafter referred to simply as ``phase space") for all 9 clusters.  The clusters are combined by normalizing each relative to its $\sigma_{v}$ and R$_{200}$ \citep[see][Table 1]{vanderburg2014}.  
\newline\indent
Figure 1 shows that there is a segregation in phase space between quiescent galaxies (red triangles) and star-forming galaxies (blue triangles).  Quiescent galaxies are typically found at smaller clustercentric radii, and lower clustercentric velocities; whereas the star-forming population is more extended in both position and velocity.  A similar segregation between these types in phase space has also been seen in lower-redshift cluster samples \citep[e.g.,][]{Carlberg1997b,Biviano2002}.  What is surprising about Figure 1 is the phase space location of the poststarburst galaxies (green stars with circles).  Similar to the quiescent galaxies, these tend to lie at smaller clustercentric radii; however, they typically have higher velocities.  Most strikingly, these galaxies seem to avoid the ``core" region in phase space, where the majority of the quiescent galaxies are located.  Indeed, they appear to form roughly a coherent ``ring" structure around the core in phase space, although some of the population does extend as far out as R $\sim$ R$_{200}$.  
\newline\indent
The phase space segregation is illustrated in the top right panel of Figure 1, where we plot the number of quiescent and poststarburst galaxies relative to the number of star-forming galaxies in three radial bins (dotted lines in the left panel), and the bottom right panel, where we plot these ratios in three phase space bins (shaded ring-shaped regions in the left panel).  The shaded regions in Figure 1 have been arbitrarily defined to enhance the contrast of the poststarbursts in phase space; however, they are similar to the variable-slope ``chevrons" that have been shown to correlate with the infall histories of galaxies in N-body simulations \citep[see e.g.,][]{Mahajan2011,Taranu2012,Oman2013}.  Within the main region of phase space they are also similar to the ``trumpet"-shaped curves of constant r/R$_{200}$ $\times$ v/$\sigma_{v}$ that have been shown to correlate with infall times by \cite{Noble2013} based on simulated accretion histories from \cite{Haines2012}.  Therefore, while the precise definition of the phase space regions is arbitrary (i.e., rings, chevrons, or curves), all three are quite similar and correlate with time since infall into the cluster, making them physically-motivated demarcations.  
\newline\indent
The right panels of Figure 1 show that the trend for the fraction of quiescent galaxies to increase towards the inner bin is roughly the same in both the phase space and radial bins.  In radial bins, the fraction of poststarburst galaxies has a peak in the middle bin; however, in phase space bins there are no poststarbursts in the core region, and the majority are confined to the middle phase space bin, with many being at low radii and high velocities.  
\newline\indent
In order to test if the three populations have distributions in phase space that are different, and that the difference is statistically significant, we perform a 2 dimensional Kolmogorov-Smirnov test of the distribution (hereafter the ``2D-KS" test).  Comparing the 2 dimensional distribution of poststarbusts to the quiescent and star-forming galaxies we find a P-values of 0.038, and 0.009, respectively.  We therefore reject the null hypothesis that they are likely to be drawn from the same distribution at $\sim$ 2$\sigma$ and 3$\sigma$, respectively.  This demonstrates that the poststarbursts have a distribution that is distinctive compared to the other galaxy types in phase space.
%\newline\indent
%We perform a likelihood-ratio test to see if the distribution of poststarbursts in the phase-space regions is consistent with being drawn from the %quiescent or star-forming galaxies.  We have adjusted the results of the likelihood test using the Bonferroni correction, which accounts for the fact %that the significance of the difference will be overestimated when comparisons are made in subsets of a dataset (i.e., corrects for the fact that %multiple arbitrarily-defined regions in phase space could have been drawn in Figure 1).  Including this correction we find probabilities of P = 5.29 $%\times$ 10$^{-6}$ and 5.46 $\times$ 10$^{-5}$ the poststarbursts could be drawn from the same distribution as the quiescent and star-forming %populations, respectively.  This demonstrates that the poststarbursts clearly have a distribution that is distinctive comported to the other galaxy %types in phase space.
\begin{figure*}
\plotone{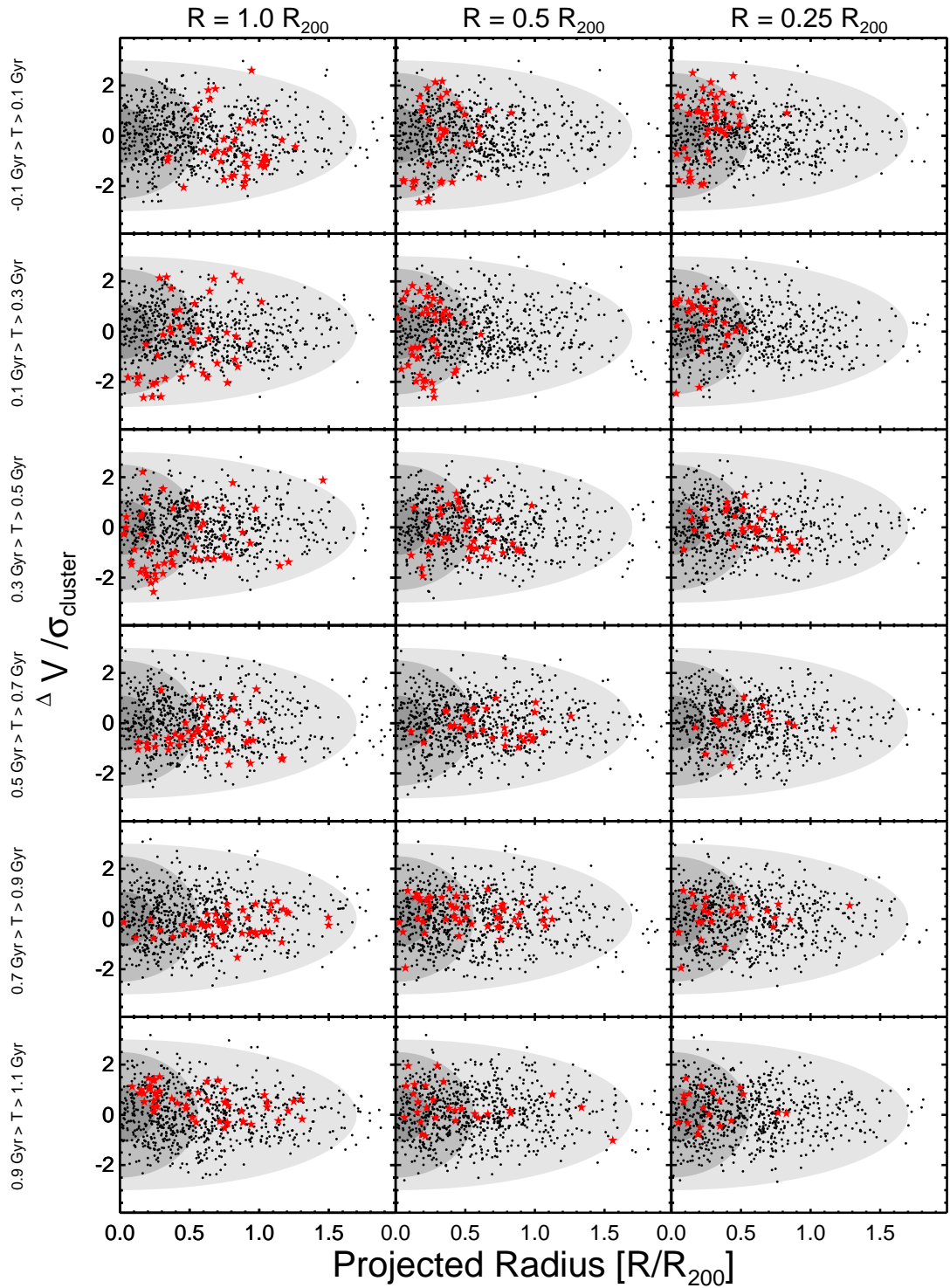}
\caption{\footnotesize The average projected phase space of subhalos in the combined dark-matter-only N-body zoom simulations of four clusters in different time steps.  The black points show all subhalos and the red stars in the three columns follow subhalos after they have made their first crossing of R = 1.0, 0.5, and 0.25 R$_{200}$.  Each row is a timestep of 0.2 Gyr after first crossing of that radius.  Similar to the observed poststarbursts (Figure 1), a ring structure can be seen in the subhalos for the shortest timesteps (T $<$ 0.5 Gyr) after subhalos have crossed the smaller radii (R $<$ 0.5R$_{200}$).  This suggests that the quenching timescale for the poststarbursts is likely to be short, and occur in the inner part of the cluster. }
\end{figure*}
\section{Simulated Cluster Phase Space}
The coherent distribution of the poststarburst galaxies in phase space suggests that it may be possible to use their distribution to constrain the location and timescale of satellite quenching.  In this section we use a set of dark-matter-only zoom simulations of clusters to test if the phase-space distribution of the poststarbursts can be described using a simple quenching model.  
\newline\indent
Our approach is to assume that satellite quenching may begin at a particular clustercentric radius.  We then follow the evolution of cluster subhalos in phase space after they make their first crossing of that radius to test if at some time step (hereafter $T$) later they resemble the phase space of the observed poststarburst galaxies.  If so, this timestep would be indicative of the satellite quenching timescale (hereafter $\tau_{Q}$).  For this simple experiment we use three possible clustercentric radii where quenching could begin: the first time a subhalo passes R $=$ 0.25, 0.5, and 1.0 R$_{200}$.  These radii roughly correspond to first passage of the cluster core, first passage of the dense intra-cluster medium (ICM), and first passage of the virial radius.  We follow the galaxies in time steps of $\Delta$$T$ = 0.2 Gyr up to 1.1 Gyr after crossing the quenching location.  This time is approximately the longest time that we can expect to detect the poststarburst signature in galaxies \citep[e.g.,][]{Balogh1999,Poggianti1999}.  
%We follow halos once they make their first crossing of several different clustercentric radii.  If at some timestep the distribution of these halos in phase space matches the observations, %then it suggests that that location may be the place where environmental quenching begins, and that the timestep may be the environmental quenching timescale.   
\newline\indent
We stress that while these models are based on cosmological simulations and therefore the orbits and infall rates may be correct, they are extremely simplified toy models for the satellite quenching process.  The satellite quenching process is almost certainly more complex, likely with multiple timescales, locations, and even a dependence on galaxy properties such as stellar mass; all of which is neglected in the current analysis.  Furthermore, some of the poststarbursts found at larger radii are almost certainly falling in to the cluster already quenched in the field, and we have not attempted to account for ``pre-quenched" galaxies.
%Furthermore, in order to model the full distribution in phase space would require including self-quenching processes, as well as a detailed %understanding of the stellar populations of objects infalling from the field at that redshift.  This is beyond the scope of the current paper.  
The goal of this modeling is not to make a comprehensive descriptive model of phase space and quenching, but simply to test if the coherent distribution of the poststarburst galaxies in phase space can be reproduced at all, and if so, what it may imply for the location and timescale for satellite quenching.  
\newline\indent
For the models we use a set of N-body zoom simulations of four clusters that were first presented in \cite{Taranu2012}.  We refer the reader to that paper for a detailed description of the simulations.  In brief, the clusters were selected for re-simulation from a larger low-resolution simulation covering 512$h^{-1}$ Mpc$^3$.  Each re-simulation has 12.5 million particles with a mass of 6.16 $\times$ 10$^8$M$_{\odot}$, and resolves Milky-Way-like halos with $\sim$ 1000 particles and Magellanic-Cloud-like halos with $\sim$ 30 particles.  The simulated clusters have virial masses of M$_{vir}$ $\sim$ 0.9 -- 1.8 $\times$ 10$^{15}$ M$_{\odot}$ at $z = 0$, and therefore are comparable to massive clusters such as Coma.  Their masses are a factor of a few smaller at $z \sim$ 1, making them an excellent match in halo mass to the GCLASS clusters.
  \newline\indent
In order to make a fair comparison between the phase space of the models and the observations, a correction for the incompleteness of the spectroscopic sample must be applied.  The GCLASS sample has a high spectroscopic completeness overall; however, there is still a spectroscopic targeting bias, with galaxies near the core being targeted more frequently, and more massive galaxies also being targeted more frequently \citep[see Figure 4 of][for the spectroscopic completeness corrections]{Muzzin2012}.  Rather than correct the observations for completeness, we have made the simulations ``incomplete" in the same way as the observations, so as to match the observed phase space.  For simplicity, the stellar mass in the simulations has been assigned by assuming a uniform stellar-mass-to-halo-mass ratio of 0.1 and the halo mass is the mass of the subhalo before it is accreted. 
\newline\indent 
In Figure 2 we plot the projected phase space of all galaxies (black points), as well as those galaxies that first crossed the three quenching radii (the three columns of Figure 2) at $T$ = 0 (red stars) and follow their evolution through various time steps.  We work in projected phase space so to match the observational data.  Generally speaking, the marked galaxies in Figure 2 follow similar orbital histories.  Most are falling directly into the cluster, likely because they are accreted through filaments in this early stage of cluster formation.  Once they pass the chosen radius they continue to pick up velocity and make a close passage of the cluster core (i.e., first pericenter).  As would be expected, the passage of the cluster core happens earlier for subhalos that we mark at smaller clustercentric radii ($\sim$ 0.1 Gyr after crossing 0.25R$_{200}$ and $\sim$ 0.4 Gyr after crossing R$_{200}$).  During their first core passage they tend to be found at high projected velocities, and small clustercentric radii.  Because of this they are not found in the central ``core" region of the cluster phase space at low velocities and small clustercentric radii where many of the massive quiescent galaxies reside in the observations (e.g., Figure 1).  Once they make a high velocity crossing of the core they tend to backsplash out to larger radii, as far out as R $\sim$ R$_{200}$.  After another $\sim$ 0.5 Gyr they begin to fall back in and appear to be much more mixed in phase space with the rest of the population.  In particular, some galaxies manage to penetrate into the low-velocity, small-radius core region in phase space.     
\newline\indent
We make a quantitative comparison between the red stars in the 18 panels of Figure 2 with the poststarbursts in Figure 1 using the 2D-KS test.  All but four of the snapshots have P $<$ 0.05 and hence we can reject the null hypothesis that they are drawn from the same distribution at $\sim$ 2$\sigma$.  The four radii/timescales that have P $>$ 0.05 and are therefore consistent with being drawn from the same distribution as the poststarbursts are for R = 0.5R$_{200}$ at $T$ = 0.0 and 0.4, and for R = R$_{200}$ at $T$ = 0.2 and 0.4.  In both cases a short timescale is favored, and quenching outside the cluster core is favored.  Interestingly, there is no long timestep from any radius where the distribution of subhalos resembles the poststarburst distribution.  This appears to rule out the possibility of long quenching times ($\tau_{Q}$ $>$ 0.5 Gyr) no matter what the quenching radius.  This may have been expected, as the poststarburst distribution appears coherent, and coherent structures will become mixed and eventually washed out in phase space on the order of a dynamical time.
\newline\indent
Formally, the quenching radii and timescales listed above are the most representative of the {\it overall} distribution of poststarbursts, which extends out to larger radii; however, from examination of Figure 2 it appears that the quenching at R = R$_{200}$ panels do not reproduce the inner ring structure of the poststarbursts particularly well.  They are likely statistically acceptable distributions because they get the overall radial distribution reasonably correct, not the ring structure.  As discussed earlier, some of the poststarburst population at larger radius is likely to have been accreted from the field ``pre-quenched".  The poststarburst fraction in the field at this redshift is $\sim$ 1 -- 3\%, \citep{Yan2009,Muzzin2012}, which for $\sim$ 400 cluster members implies a maximum of 4 - 12 galaxies may be infalling already quenched.  This number is of order the total number of poststarbursts seen at R $>$ 0.5 R$_{200}$ (see Figure 1).  If the poststarbursts are larger clustercentric radii are pre-quenched, then it may be that the inner ring of poststarbursts is the primarily signature of satellite quenching, and is the distribution that should be matched in the simulation.
\begin{figure*} 
\plotone{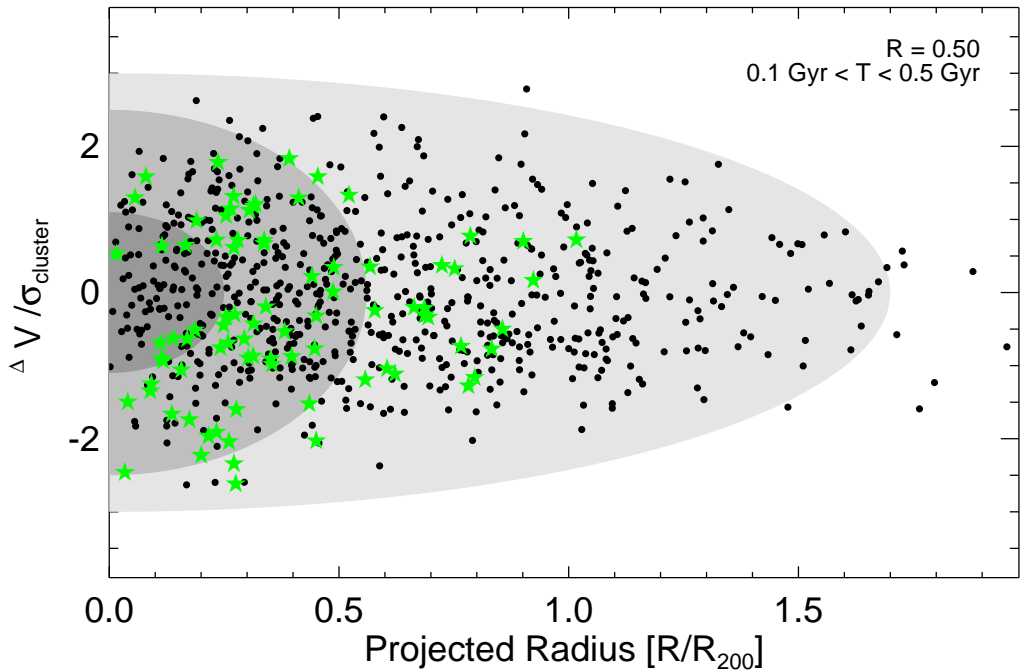}
\caption{\footnotesize The phase space of the simulations with the same phase space contours as in Figure 1.  Green stars show galaxies that first crossed R = 0.5R$_{200}$ within the last 0.1 $< T <$ 0.5 Gyr.  This is the only quenching timescale and location that passes a 2D-KS test for both in the inner and outer regions and therefore have a distribution that is consistent with having been drawn from the poststarburst distribution in Figure 1 }
\end{figure*}  
\newline\indent
Given this, we used the 2D-KS test to determine in which timesteps the distribution of subhalos is consistent with the distribution of the poststarburst population using only subhalos and galaxies at R $<$ 0.5R$_{200}$, where the ring structure is strongest.  These results are quite different from the overall 2D-KS test.  The two timesteps for R = R$_{200}$ that were favored by the full 2D-KS test have much smaller P values when only galaxies at R $<$ 0.5R$_{200}$ are considered (P = 0.06 and 0.05), and are rejected at the $\sim$ 2$\sigma$ level.  The only timesteps that are consistent with the observations are for quenching at R = 0.5R$_{200}$ at $T$ = 0.1, 0.3, and 0.5, and for R = 0.25R$_{200}$ at $T$ = 0.1.  The strong ring structure from these panels can also be seen quite clearly by eye in Figure 2.  The reason some of these are not formally the good descriptions of the data in the full 2D-KS test is simply because they fail to reproduce the few poststarburst galaxies at larger radii that are seen in the observations.  Therefore, this may actually make these the most likely candidates for the location and timescale of the dominant satellite quenching process, even though several of them fail to reproduce the full phase space distribution of phase space.
%This may not rule them out as possible candidates for the quenching location and timescale because some of the galaxies at larger radii are almost certainly infalling from the field already quenched. The %poststarburst fraction in the field at this redshift is $\sim$ 1 -- 3\%, \citep{Yan2009,Muzzin2012}, which for $\sim$ 400 cluster members implies a maximum of 4 - 12 galaxies may be infalling already %quenched.  Given the rarity of poststarbursts at larger radii, a pre-quenched field population could account for most of them.  If so, then the quenching timescales and locations that better reproduce %the ring structure could be acceptable, even if they do not reproduce the full distribution of poststarbursts well.
\newline\indent
Ignoring this possibility for the moment, and attempting match both the inner ring and the overall distribution of the poststarbursts with a single model, we combined a few of the timescales to produce the best-possible description of the observations.  This is shown in Figure 3 where we plot the distribution of galaxies quenched at R = 0.5R$_{200}$ on a timescale of 0.1 $< T <$ 0.5.  A 2D-KS test applied to that dataset provides values of P = 0.358 and 0.185 for all galaxies, and galaxies at R $<$ 0.5R$_{200}$, respectively, and hence is the only model that has acceptable P-values for both the overall distribution and the inner ring structure.  We therefore adopt it as our best overall model for the location and timescale of satellite quenching.
\begin{figure*}
\plotone{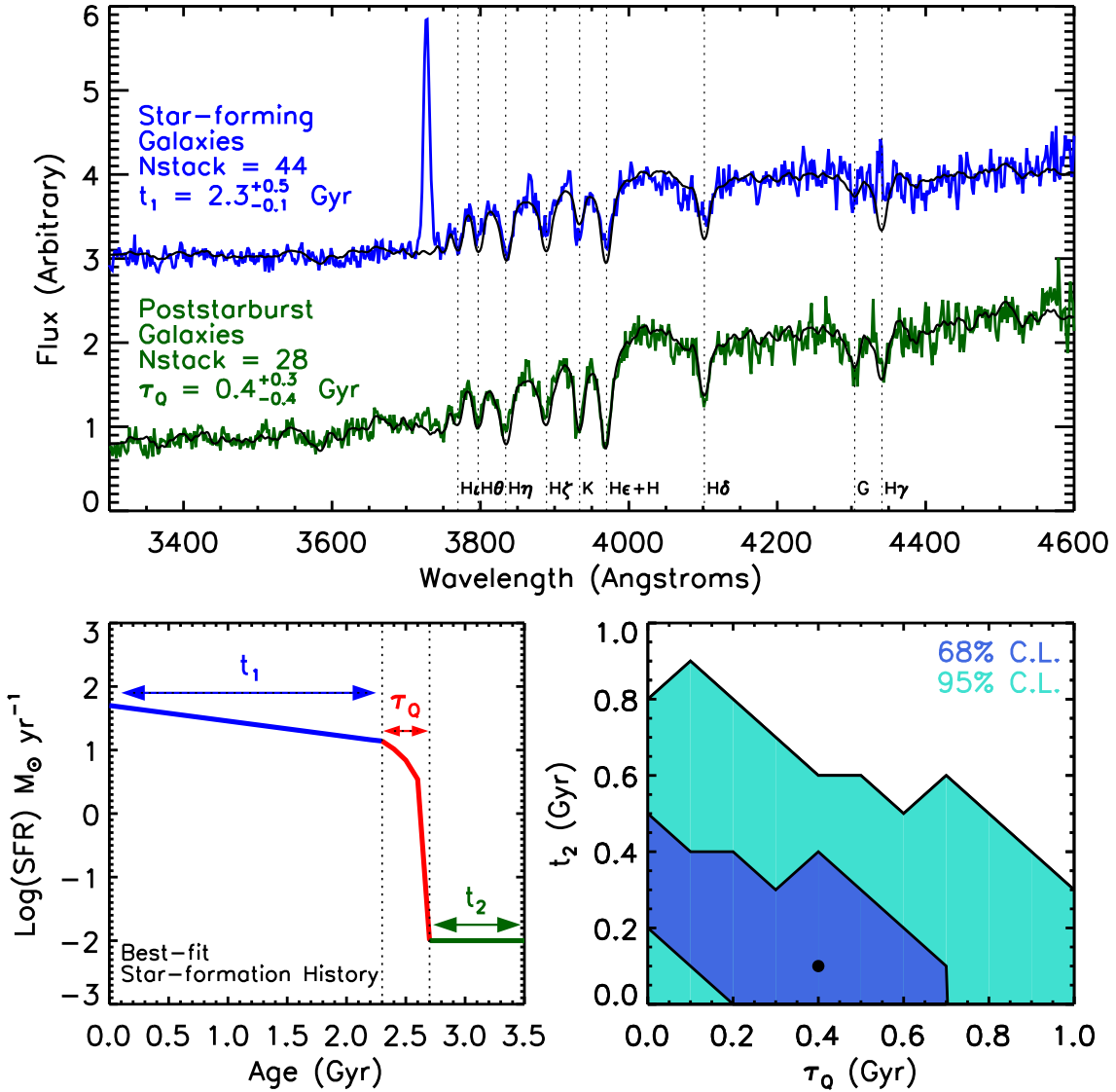}
\caption{\footnotesize Top panel:  Mean stacked spectrum of star-forming (blue) and poststarburst (green) galaxies in the GCLASS clusters with prominent absorption features labelled.  The overlaid black spectra show the best-fit \cite{Bruzual2003} spectrum for these types assuming the star-formation history in the lower left panel.  The most constraining feature in the star-forming spectrum is the Calcium K line, which implies the population is somewhat evolved.  The most constraining features in the poststarburst spectrum are Calcium K and the G-band which also imply a more evolved population, but also the deep H$\delta$ absorption which implies a recent end to the star-formation.  Bottom left panel: The best-fit star formation history assuming the poststarburst galaxies are descendants of the star-forming galaxies and undergo a quenching process.  Bottom right panel: Confidence intervals on the quenching timescale ($\tau_{Q}$) and time since star formation ended ($t_{2}$).  These timescales are consistent with those derived from the phase space analysis.}
\end{figure*}
\section{SED fitting of Poststarburst Galaxies}
The phase space modeling suggests that satellite quenching occurs on a timescale of roughly 0.1 Gyr $< \tau_{Q} <$ 0.5 Gyr after galaxies make their first passage of R $\sim$ 0.5R$_{200}$.  The simulations imply that the median time it takes for an accreted galaxy to travel from 2.0R$_{200}$ to 0.5R$_{200}$ is $\sim$ 0.7 $\pm$ 0.2 Gyr, where the uncertainty is the 1$\sigma$ rms dispersion in infall times.  This scenario is qualitatively similar to the ``delayed-then-rapid" quenching model proposed by \cite{Wetzel2013}, where the quenching time appears to be roughly similar, but the delay time may be a factor of $\sim$ 2 -- 4 faster at $z \sim$ 1.  We note that this comparison of delay time is qualitative, because our delay time is inferred since first passage of 2.0R$_{200}$, whereas the \cite{Wetzel2013} delay time is measured as the time since a galaxy is identified as a subhalo of a parent halo in an N-body simulation using friends-of-friends linking.  It is therefore not immediately clear how these two definitions are related.  Although qualitative, it is unlikely that the longer end of the delay time proposed by \cite{Wetzel2013} (e.g., 3 -- 4 Gyr) could be supported by the data at $z \sim$ 1 for two reasons.  Firstly, because the simulations show that almost all subhalos would have already made a passage of R = 0.5R$_{200}$ and hence should be quenched based on our model, and secondly, because this long delay time is approaching the age of the universe at $z =$ 1 ($\sim$ 6 Gyr) which would require clusters to form at very high redshift (z $>$ 3).
\newline\indent
If the timescales in our phase space model are correct, they place constraints on the ages of the stellar populations of both the poststarburst galaxies and their star-forming progenitors.  Poststarburst galaxies should have star formation histories that are consistent with this rapid quenching timescale, and cluster star-forming galaxies should be at least old enough that they have had time to migrate from the cluster outskirts to R = 0.5R$_{200}$ while maintaing active star formation.
\newline\indent
In order to test the consistency of the phase space model and the stellar populations of the galaxies, we fit the spectra of both the star-forming galaxies and the poststarbursts, with the assumption that the cluster star forming galaxies are the progenitor population of the poststarburst population.  Our methodology is as follows: we fit the age of the star-forming galaxies to define the initial age of the poststarburst galaxies once quenching began, defined as $t_{1}$.  We then fit the poststarburst spectrum starting from this age and fit for two parameters, $\tau_{Q}$, the timescale over which quenching occurred, and $t_{2}$, the age of the galaxy since it was fully quenched.  This fiducial star-formation history (SFH) is illustrated in the bottom left panel of Figure 4, and is, again, schematically similar to the ``delayed-then-rapid" SFH proposed by \cite[][see their Figure 12]{Wetzel2013}
\newline\indent
We begin by stacking the spectra of all star-forming galaxies \citep[details of the stacking process are discussed in][]{Muzzin2012}, with an additional cut requiring that D$_{n}$(4000) $<$ 1.45.  There are relatively few star-forming galaxies with D$_{n}$(4000) $>$ 1.45; however, given that this cut is a requirement for the poststarburst selection, it is impossible that star forming galaxies with D$_{n}$(4000) $>$ 1.45 can be the progenitors of the poststarburst population.  The top panel of Figure 4 shows the mean stacked spectrum of the star-forming galaxies with prominent absorption features labelled.
\newline\indent
To fit the spectrum we employ the high-resolution models from \cite{Bruzual2003} assuming solar metallicity, a \cite{Calzetti2000} dust law (A$_{v}$ = 0 -- 4), and a \cite{Chabrier2003} IMF.  We degrade the spectral resolution of the models to match that of the data which is 17\AA~in the observed frame, and corresponds to $\sim$ 9\AA~rest-frame).  We assume a declining SFH based on the decline of the global star-formation rate (SFR) from $z =$ 3 to $z =$ 0 compiled by \cite{Bouwens2012}.  We also tried a SFH with a constant SFR; however, this continued high level of star formation cannot reproduce the strength of the Calcium K absorption line in the star-forming galaxies even at old ages and therefore some form of a declining SFH is required.  The \cite{Bruzual2003} models do not contain emission lines, so the [OII] emission line is not fit by the model.  
\newline\indent
The black solid line in Figure 4 shows the best-fit model to the star-forming population which has a time since star-formation began of t$_{1}$ = 2.3$^{+0.5}_{-0.1}$ Gyr.  This implies a mass-weighted age of the stellar population of 1.5$^{+0.2}_{-0.1}$ Gyr.  The strongest constraint on the age comes from the relatively strong Calcium K line, which cannot be reproduced by a very young stellar population.  This intermediate-age population is old enough that it would be continuously forming stars over the infall time from R = 2.0R$_{200}$ to R = 0.5R$_{200}$, and therefore is consistent with the delay time implied by the simulation.
\newline\indent
To fit the poststarburst population, we create a set of new model grids with the same range of ages and dust extinctions as the declining SFH of the star-forming galaxies.  We then add a linear quenching of star formation at $t_{1}$ = 2.3 Gyr, and create 11 new grids with $\tau_{Q}$ ranging from 0 - 1.0 Gyr in steps of 0.1 Gyr, where $\tau_{Q}$ is the time between when star formation starts to decline and SFR = 0 (see Figure 4).  We then fit the poststarbust spectrum to each of these grids and fit for t$_{2}$ (the time since SFR = 0) and A$_{v}$.  We then find the minimum $\chi^2$ across the 11 grids which gives us a best fit $\tau_{Q}$, t$_{2}$, and A$_{v}$. 
\newline\indent
The best fit model for the SFH of the poststarbursts (t$_{1}$ = 2.3$^{+0.5}_{-0.1}$ Gyr, $\tau_{Q}$ = 0.4$^{+0.3}_{-0.4}$ Gyr, t$_{2}$ = 0.1$^{+0.4}_{-0.1}$ Gyr) is plotted in Figure 4 as the black model on top of the green spectrum.  The best fit reproduces the line strengths of the Calcium K line, the G-band, and H$_{\delta}$ impressively well.  This is remarkable in the sense that Calcium K and the G-band are typical of older stellar populations, whereas strong H$_{\delta}$ is from recently-quenched galaxies.  It would be difficult to reproduce such a spectrum with any form of a SFH other than that in Figure 4.  
\newline\indent
Remarkably, the implied SFH, $\tau_{Q}$, and best-fit ages of the stellar populations inferred from the spectral fitting are consistent with the timescales implied by the phase space modeling.  It is possible that the SFH and ages could be fit by a wider range of models which we have not explored in exhaustive detail; however, the consistency between the two with basic modeling is encouraging.  It is particularly encouraging considering that the stellar population measurements of the satellite quenching timescale are completely independent of the velocities and positions of the galaxies within the cluster.  
\section{Discussion}
%The satellite quenching scenario at $z \sim$ 1 suggested by the phase space modeling and the stellar populations is consistent with the ``delayed-%then-rapid" quenching proposed by \cite{Wetzel2013}, albeit with a delay time that could be a factor of 2 -- 4 faster, and a quenching time that could %be as much as a factor of 2 faster, but is formally consistent with being the similar.  
%\newline\indent
The overall picture of satellite quenching that arises from the modeling is that star-forming galaxies at $z \sim$ 1 are not immediately quenched once they are accreted by a cluster/group.  Instead, they evolve as normal star-forming galaxies as they fall into the central regions of the cluster.  Based on the N-body simulation, it takes $\sim$ 0.7 Gyr for a galaxy to first cross R $\sim$ 0.5R$_{200}$, and this is where satellite quenching begins.  Once the satellite quenching process starts, it proceeds on a short timescale 0.1 $<$ $\tau_{Q}$ $<$ 0.5 Gyr.  Moreover, not only is the quenching timescale short, but most of the observed poststarbursts are consistent with having their star formation fully ended only recently.  This observation of most poststarbursts having been recently quenched is not a selection effect, as our selection criteria for the poststarbursts of D$_{n}$(4000) $<$ 1.45 and [OII] $<$ 3\AA~could select galaxies as old as $\sim$ 1.0 - 1.5 Gyr, depending on the $\tau_{Q}$.
\newline\indent
This scenario and the inferred timescales are remarkably similar to that presented by \cite{Smith2010} for the Coma cluster.  They found that galaxies with asymmetric UV morphologies were concentrated within the central 500 kpc (i.e., R $\sim$ 0.25 R$_{200}$) of the Coma cluster.  With numerical modeling they showed that this population could be explained if they were quenched at R $\sim$ 1 Mpc (i.e., R $\sim$ 0.5 R$_{200}$) and they were viewable for $\sim$ 0.5 Gyr after quenching began.  This quenching location and timescale are nearly identical to that derived from the phase space analysis at $z \sim$ 1.
\newline\indent
Our inferred timescale is also consistent with the quenching timescale at $z \sim$ 1 proposed by \cite{Mok2013} based on their analysis of the fraction of red, blue, and ``green" galaxies in groups.  They showed that in order to properly reproduce the fractions of these galaxy types, delay times before quenching begins of $<$ 2 Gyr were required along with a $\tau_{Q}$ $<$ 1 Gyr.  Our quenching timescale is also similar to the quenching timescale proposed by \cite{Wetzel2013} at $z =$ 0 ($\tau_{Q}$ = 0.2 -- 0.8 Gyr); however, it appears that the delay time before quenching may be a factor of 2 -- 4 longer at $z =$ 0.   
\newline\indent
One potential issue with the proposed scenario for satellite quenching is that we have pre-selected only poststarburst galaxies as our tracer of quenched/quenching population.  If there are galaxies that quench on very long timescales, they will have been omitted from this analysis.  While we cannot formally rule this out, we note that a significant number of studies have shown that the SSFR of star-forming galaxies has little dependence on clustercentric radius at both high and low redshift \citep[e.g.,][]{Patel2009,Vulcani2010,Muzzin2012,Wetzel2013}.  Although this does not rule out some population of slow quenching satellite galaxies, it strongly suggests that slow quenching cannot be the dominant type of quenching for satellites.   
\newline\indent
It is remarkable that the quenching timescales determined from four independent methods \citep[][and the current phase space analysis]{Smith2010,Wetzel2013,Mok2013} are consistent. The phase space constraints are particularly useful because they also provide an additional constraint on where within the cluster/group the satellite quenching begins.  With both an inferred location and timescale, we can attempt to infer the physical process that may be responsible.  
\newline\indent
We first consider the location within the cluster where quenching begins.  As derived by \cite{Treu2003} and \cite{Moran2007}, quenching from mergers or high-speed galaxy interactions (``harassment") occurs preferentially at R $>$ R$_{200}$, which is inconsistent with our derived quenching radius.  Likewise, tidal processes such as halo stripping or disruption occur close to the cluster core (R $<$ 0.25 R$_{200}$), and also seem inconsistent with our data, where the implied quenching radius is R $\sim$ 0.5R$_{200}$.  Hot halo gas stripping (``strangulation") and cold gas stripping (``ram-pressure") are most effective where the ICM is dense, at roughly R $<$ 2.0R$_{200}$ and R $<$ 0.5R$_{200}$, respectively \citep[e.g.,][]{Bahe2013}.  Based on the location where we expect quenching to occur, these appear to be the best candidates for the physical process.
\newline\indent
If the timescale for quenching is considered at face value, it would suggest that complete removal of the galaxy cold gas via ram-pressure stripping will be necessary to quench galaxies so rapidly. However, \cite{Carilli2013} have compiled the latest measurements of gas fractions (i.e., f$_{gas}$ = M$_{gas}$/(M$_{gas}$+M$_{star}$)) of galaxies at $z \sim$ 1 and these are of order 0.3.  Our measurements of the SSFRs of galaxies in the GCLASS sample \citep[see][]{Muzzin2012} show that galaxies with stellar masses of a few times 10$^{10}$ M$_{\odot}$ (the typical stellar mass of the poststarburst galaxies) have Log(SSFRs) $\sim$ -8.8, which implies that if they were cut off from their hot gas supply completely, they would consume their cold gas in $\sim$ 0.5 Gyr.  This is consistent with the long end of quenching timescale that we derive; and therefore it means that we cannot formally rule out hot gas stripping as a plausible mechanism for quenching the poststarbursts in the cluster.
\newline\indent
It is also worth considering that the cold gas consumption timescales tend to evolve to longer values at lower redshift.  This is simply because SSFRs decline faster than f$_{gas}$ with decreasing redshift \citep[e.g.,][]{Carilli2013}.  This evolution of the cold gas consumption timescale is fairly weak (evolving from $\sim$ 0.5 Gyr at $z \sim$ 1, to $\sim$ 1 Gyr at $z \sim$ 0); however, it is potentially measurable.  If cold gas stripping (i.e., ram-pressure) is the dominant satellite quenching process, we might expect that the satellite quenching timescale may not evolve with redshift.  However, if hot gas stripping dominates then the evolution of the cold gas consumption timescale would suggest that the satellite quenching timescale may also evolve to longer values at lower redshift.  Therefore, measuring the redshift evolution of the satellite quenching timescale may be a useful approach for identifying whether hot or cold gas stripping is the dominant physical process for satellite quenching.  Interestingly, lower redshift studies such as \cite{Wetzel2013} seem to suggest slightly longer quenching timescales ($\tau_{Q}$ = 0.2 -- 0.8 Gyr), although this has significant overlap with the timescale derived at $z \sim$ 1 ($\tau_{Q}$ = 0.1 -- 0.5 Gyr).  If this could be shown to be statistically significant, then it would imply that hot gas stripping is likely the most important satellite quenching process.   However, given the large uncertainties at present this clearly requires more detailed investigation.  It would also benefit from a comparison of timescales measured using similar techniques to avoid systematic errors.
\section{Summary and Outlook}
In this paper we have shown that the population of poststarburst galaxies in clusters at $z \sim$ 1 has a distribution in phase space that is distinctive from both the quiescent and star-forming cluster galaxy populations.  Using a set of dark-matter-only zoom simulations of clusters we showed that this distribution can be recovered if galaxies quench on a rapid timescale (0.1 $<$ $\tau_{Q}$ $<$ 0.5 Gyr) after first passage of R $\sim$ 0.5R$_{200}$.  The simulations also show that longer quenching timescales ($\tau_{Q}$ $>$ 0.5 Gyr), or quenching at R $\sim$ R$_{200}$ provide poor descriptions of the poststarburst phase space distribution.  Fitting of the stacked spectra of the star-forming and poststarburst galaxies shows that the SFH, $\tau_{Q}$, and ages of their stellar populations are consistent with the timescales derived from the phase space analysis, and the similarity between these independent indicators provides the strongest consistency check of the overall model.  
\newline\indent
The derived quenching location and timescale suggest that gas stripping processes are most likely responsible for quenching the satellite population; however, the current constraints are not strong enough to distinguish between hot or cold gas stripping as the dominant quenching mechanism.  Measurement of the evolution of the quenching timescale and possibly the location could be extremely valuable for determining the dominant physical process.  Cluster samples with high-quality spectroscopic data exist at lower redshift, and so this is a tractable problem for the future. 
\newline\indent
An additional conclusion of this work is that the approach of using phase space and simulations to constrain the location and timescale of satellite quenching seems to be a promising new way forward on this problem \citep[see also][]{Noble2013}.  We note that better constraints at $z \sim$ 1 using this approach could be made with larger samples of spectroscopic cluster members.  The results of most of the 2D-KS tests in this work provide constraints on the timescales at the $\sim$ 2$\sigma$ level, with the limitation being the total number of poststarbursts in the sample (only 28 galaxies out of 424 cluster members).  Increasing the number of spectra of poststarbursts by a factor of a few would allow constraints at $\sim$ 3$\sigma$ level or better, which would be useful for further refining the quenching model.  Also, more detailed modeling of the infalling process that includes tracking self-quenching of the infalling field population would be useful for understanding the population of poststarbursts are larger radii and putting tighter constraints on the quenching timescale.  This will be addressed in future papers.
\acknowledgements
We acknowledge the work of Chris Power and Brad Krane in developing the N-body simulations used in this analysis.  AM and MF acknowledge support from an NWO Spinoza grant.  RFJvdB and HH acknowledge support from the Netherlands Organisation for Scientic Research grant number 639.042.814.  MLB acknowledges support from an NSERC Discovery Grant, and from NOVA and NWO visitor grants that supported his sabbatical at Leiden Observatory, where this work was completed. MJH acknowledges support from an NSERC Discovery Grant.

\bibliographystyle{apj}
\bibliography{apj-jour,myrefs}

%\begin{thebibliography}{}

%\bibitem{Bru03} Bruzual, G., \& Charlot, S. 2003, MNRAS, 344, 1000

%\end{thebibliography}

\end{document}